\documentclass[aps,pre,twocolumn,superscriptaddress,groupedaddress]{revtex4}
\usepackage{hyperref}
\usepackage{color}
\usepackage{graphics,graphicx,epsfig,wrapfig}
\usepackage{epsf,epstopdf}
\usepackage{amssymb,amsfonts,amsmath,bbold}
\usepackage{enumitem}

\usepackage{ifthen}

\newcommand\mydots{\hbox to 0.8em{.\hss.\hss.}}
\newcommand{\beqn}{\begin{eqnarray}}
\newcommand{\eeqn}{\end{eqnarray}}
\newcommand{\beq}{\begin{equation}}
\newcommand{\eeq}{\end{equation}}

\newcommand{\<}{\langle}
\renewcommand{\>}{\rangle}

\definecolor{orange}{RGB}{192,95,0}
\definecolor{darkgreen}{RGB}{0,150,100}

\begin{document}
\title{Population dynamics of immune repertoires}

\author{Jonathan Desponds}
\affiliation{University of California San Diego, Department of Physics, La Jolla, CA 92093, USA}
\author{Andreas Mayer}
\affiliation{Laboratoire de physique th\'eorique,
    CNRS, UPMC and \'Ecole normale sup\'erieure, 24, rue Lhomond,
    75005 Paris, France}
\author{Thierry Mora}
\affiliation{Laboratoire de physique statistique, CNRS, UPMC and \'Ecole normale sup\'erieure, 24, rue Lhomond, 75005 Paris, France}
\author{Aleksandra M. Walczak}
\affiliation{Laboratoire de physique th\'eorique,
    CNRS, UPMC and \'Ecole normale sup\'erieure, 24, rue Lhomond,
    75005 Paris, France}

\date{\today}
\linespread{1}

\begin{abstract}
The evolution of the adaptive immune system is characterized by changes in the relative abundances of the B- and T-cell clones that make up its repertoires. To fully capture this evolution, we need to describe the complex dynamics of the response to pathogenic and self-antigenic stimulations, as well as the statistics of novel lymphocyte receptors introduced throughout life. Recent experiments, ranging from high-throughput immune repertoire sequencing to quantification of the response to specific antigens, can help us characterize the effective dynamics of the immune response. Here we describe mathematical models informed by experiments that lead to a picture of clonal competition in a highly stochastic context. We discuss how different types of competition, noise and selection shape the observed clone-size distributions, and contrast them with predictions of a neutral theory of clonal evolution. These mathematical models show that memory and effector immune repertoire evolution is far from neutral, and is driven by the history of the pathogenic environment, while naive repertoire dynamics are consistent with neutral theory and competition in a fixed antigenic environment. Lastly, we investigate the effect of long-term clonal selection on repertoire aging. 
\end{abstract}

\maketitle
 \section{Introduction}
 \label{secintro}
 
B-cells and T-cells specificity is mediated by antigen-recognition receptors located on their surfaces, which are unique to each cell. B- and T-cells form the part of the immune system that is called adaptive, because the abundance of cells expressing each receptor type can be modulated to meet the challenges of the antigenic environment. T-cell receptors (TCR) and B-cell receptors (BCR) are produced through a random process of gene editing called VDJ recombination. {Receptors are passed on to offspring upon division (unchanged in T-cells, and altered by somatic hypermutations in B-cells)} \cite{Janeway}. Cells that share a common receptor define a clone. The set of clones that a body possesses constitutes its immune repertoire. This repertoire must be diverse enough to face any potential pathogenic invasion, and precise enough to react quickly to any threat.
 
The lymphocyte population evolves by division, differentiation and death signals that are mainly of two types: antigenic and hormonal, with the antigenic signals being specific to the clone and its receptor. Large scale divisions that give rise to memory and effector cells are triggered by pathogenic antigens, while naive cells require short binding events to self-antigens to survive \cite{Troy2003, mak2006immune}. The evolution of the immune system can be understood at two very different time scales: over evolutionary timescales, with the shaping of the mechanisms of immunity through natural selection; and over the lifetime of an individual, through competition of immune cells for antigen and cytokines. In this chapter we focus on the latter.

The high-throughput sequencing revolution of the last decade has allowed for the deep sequencing of {BCR and TCR repertoires} \cite{Weinstein2009,Boyd2009a,Robins2009,friedman-2012, chain-2014, greenberg-2012, robins-2011, mamedov-2014, Warren2011}. The diversity and distribution of receptor sequences are the results of repertoire evolution. {Statistical features of immune repertoires can thus} be used as a way to probe the rules that govern its dynamics. Massive receptor sequence data has been used to characterise the mechanisms of receptor generation and selection \cite{murugan-2012,walczak-2014}, and the hypermutation process in B cells \cite{Yaari2013a,Elhanati2015,Mccoy2015}.
Another way to analyse repertoires is to count the number of times each unique sequence appears. Unique molecular barcoding now allows {us to obtain reliable} counts of receptor mRNA molecules through correction of sequencing errors and PCR amplification noise \cite{Vollmers2013,Best2015b,Shugay2014a}.
Using abundance information can be useful in the clinic, e.g. for tracking clone sizes upon vaccination \cite{Vollmers2013,Laserson2014,chain-2014,Galson2014} or in leukemia patients \cite{Wu2012,Salson2016}.

\begin{figure}
\begin{center}
\includegraphics[width=\linewidth]{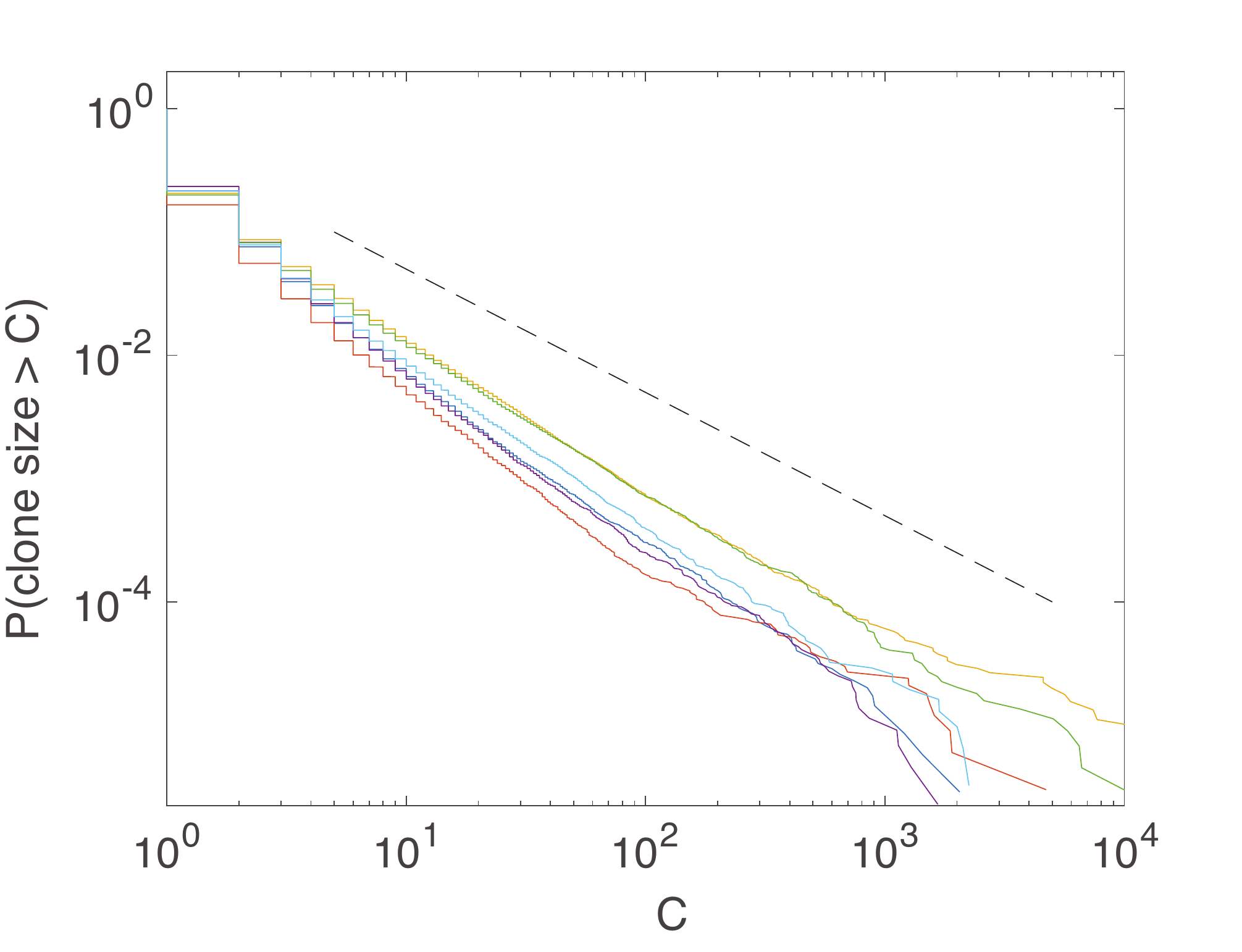}
\caption{Cumulative clone size distributions of unsorted human T cells from \cite{Pogorelyy2016,Mora2016e} follow a power law over several decades. Each colour is one individual. Clone sizes correspond to the number of distinct molecular barcodes associated with each nucleotide sequence of the beta chain (TRB).}
\label{fig1}
\end{center}
\end{figure}

Using sequence counts, one can gather all clones of similar size to form the clone size distribution in the repertoires of healthy individuals. These statistics, unlike the abundances of particular receptor clones, are fairly robust to sampling noise, as well as to individual-to-individual variability.
Clone size distributions of unsorted repertoires and of subsets of effector and memory cells have been reported to be heavy-tailed or even to follow a power law \cite{Weinstein2009,Mora2010,Zarnitsyna2013,Bolkhovskaya2014a,Menzel2014,Muraro2014,Pogorelyy2016,Mora2016e,deBoerChain}. An example of these distributions is given in Fig.~\ref{fig1}. Note that similar behaviour has not been established for naive cells. These long tails in the clone-size distribution put strong constraints on the class of mathematical models that one could propose for the repertoire dynamics. In addition, the quantitative features of the clone-size distribution can in principle be used to extract information about the nature and scale of the dynamical processes at play.

{Much of the early mathematical modeling work on adaptive immunity has focused} on the response of single clones or subsets of clones to antigenic challenges, using ordinary differential equations \cite{Perelson2002}.
{More recently interest has shifted to understanding the global dynamics of lymphocyte populations at the level of the repertoire \cite{Johnson2012,Mayer2015,Lythe2016}.}
To model the repertoire as a population of clones, one can find inspiration in ecology or population genetics. There is however an important difference between the adaptive immune system and classic models of evolution such as Kimura's neutral model \cite{kimurabook} or the selective sweeps model \cite{Nielsen2005}. In the immune system, sequence diversity is produced de novo through VDJ recombination in the thymus and in the bone marrow, rather than from mutations (with the exception of hypermutations of BCR in germinal centers). It is necessary to develop new tools to tackle these challenges. {In this paper we give a unified overview of recent models that attempt to describe the somatic population dynamics in the adaptive immune system. We revisit previously studied models in terms of the clone size distributions they predict, and investigate aging of naive repertoires using a minimal tractable model.}

\maketitle
 \section{General model}
 \label{genmod}
One of the first models of population dynamics in the adaptive immune system was introduced by de Boer and Perelson and developed in a series of papers which explicitly models competition between clones for antigenic resources \cite{DeBoer1994,DeBoer1995,DeBoer1997,DeBoer2001}.
More recent works have analysed very similar models \cite{Lythe2016,desponds2016}. All these previous models can be encompassed within a common mathematical framework, which we describe now.

The general idea behind this class of models is that one important signal for which lymphocytes compete comes from antigens. Strong antigenic recognition by mature lymphocytes generally triggers clonal expansion into effector and memory cells. These strong signals are usually of pathogenic origin with the exception of autoimmune reactions.
Antigens produced by the self usually do not trigger large scale proliferation or differentiation, as peripheral cells have been selected against auto-immunity. They can however provide naive cells with survival cues or control their homeostatic proliferation.
In sum, the ability of each cell to bind different peptides determines its propensity to divide and die, which in turn depend on antigenic signals. In that description, all cells of a clone have the same division and death rates, although more elaborate models can include fluctuations in the state of each cell through numbers of surface receptors, expression of genes, or concentrations of signaling molecules (we discuss this extension in Section~\ref{nonspec}).

In mathematical terms, the T-cell or B-cell repertoire is described by a set of $N$ clones with abundances $C_i(t)$, $1\leq i \leq N$, and the environment by a set of $M$ antigenic peptides with concentrations $a_j(t)$, $1\leq j \leq M$. The binding probabilities between antigens and clones are encoded in an $N\times M$ interaction matrix $K$, where $K_{ij}$ is the probability for antigen $j$ to bind receptor $i$.
The dynamics of each clone are governed by division and death, which occur with Poisson rates that depend on a receptor-specific antigenic stimulus, $S_i$, which we will specify later. We denote by $\nu(S_i)$ the division rate (an increasing function of $S_i$ to model antigen-driven proliferation) and by $\mu(S_i)$ the death rate (a decreasing function of $S_i$ to model survival signals). Clone sizes $C_i$ follow continuous Markovian dynamics, with transitions:
\beq
\left\{
    \begin{array}{ll}
       C_i\to C_i+1 & \textrm{with rate } \nu(S_i) C_i,\\
        C_i \to C_i-1  & \textrm{with rate } \mu(S_i) C_i.
    \end{array}
\right. 
\label{disc}
\eeq
When considering large clones, the stochastic nature of division and death is often neglected, yielding a continuous version of Eq.~\ref{disc},
\beq
 \partial_t C_i = [\nu(S_i) - \mu(S_i)] C_i.
 \label{gendyn}
\eeq
The clone-specific stimulus is defined as the sum of the stimuli provided by all antigens:
\beq\label{stim}
S_i= \sum_{j=1}^M K_{ij} F_ja_j ,
\eeq
where $F_j$ is an antigen-specific factor that quantifies its availability, and thus models competition: the more clones are specific to antigen $j$, the less available it will be.
$F_j$ can take many forms, but a simple one that is consistent with most previously proposed models is:
 \beq
 F_j = \frac{1+\epsilon}{1+\epsilon \sum_{i=1}^N K_{ij} C_i},
 \label{ava}
 \eeq
where the parameter $\epsilon$ sets the strength of competition.
 One can view antigens as resources that mediate survival or growth. Competition for these antigenic resources ensures good coverage of antigenic space by immune repertoires, consistent with the observed efficiency of adaptive immune systems \cite{Mayer2015}.

Eqs.~\ref{disc} or \ref{gendyn} define the dynamics of particular clones, but the number and identities of clones themselves may fluctuate. New clones are introduced into the system through thymic (T-cell) and bone marrow (B-cell) output, with a Poissonian rate $\theta$, and an introduction size drawn at random from a distribution $P_0(C)$. Clones
go extinct when $C_i$ reaches 0. The balance between the introduction and extinction of clones allows for the existence of a steady state. 
The entries of $K$ are usually drawn at random from a specified distribution, the precise choice of which is somewhat arbitrary and usually not crucial, as we will discuss.
 
 \section{Neutral theory}
 \label{neutsec}
 
The stochastic nature of division and death in Eq.~\ref{disc} leads to fluctuations that are known as demographic or birth-death noise. Before considering the effect of competition on diversity and the distribution of clone sizes, it is interesting to first consider the effect of this demographic noise on the population with constant birth and death rates $\nu<\mu$. This situation corresponds to Kimura's neutral model of evolution \cite{kimurabook}, in which new variants are generated by VDJ recombination with rate $\theta$ and an introduction size drawn from $P_0(C)$.
 
\begin{figure}
\begin{center}
\includegraphics[width=\linewidth]{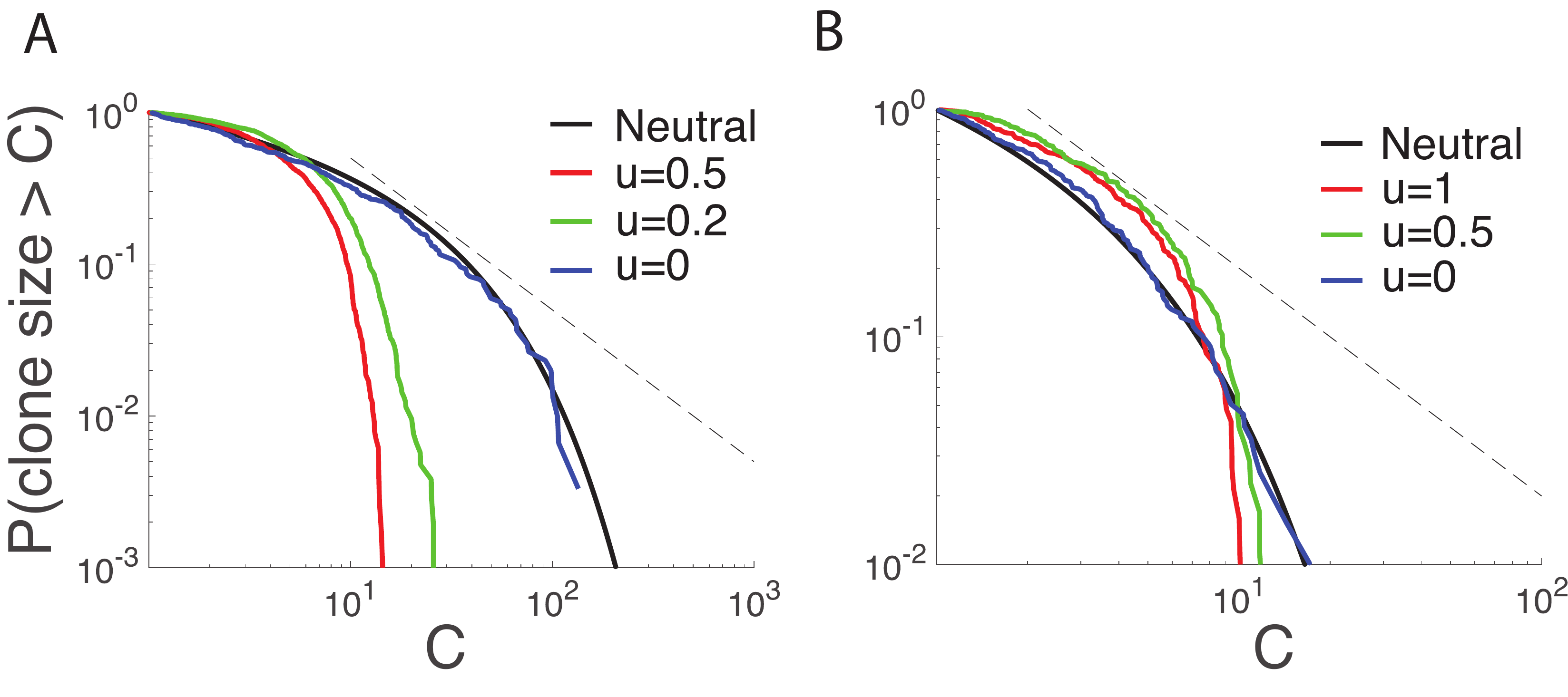}
    \caption{Competition in neutral environments limits the size of large clones relative to neutral theory. Cumulative clone size distributions are shown after $2000$ days of simulation of the dynamics of Eqs.~\ref{gendyn}-\ref{ava} with $\epsilon=4$, constant $\nu$  and $\mu(S_i)=\mu_0/(1-u+uS_i)$, where $u$ controls how much the antigen environment affects clone dynamics. There are $M=100$ antigens with constant concentrations $a_j=1$; entries of the $K$ matrix are $1$ with probability $p$ and $0$ otherwise. The initial size of new clones is 10, $P_0(C)=\delta_{C,10}$. Demographic noise is added to the simulation of Eq.~\ref{gendyn} as a Gaussian white noise of amplitude $\sqrt{(\nu+\mu(S_i))C_i}$. The black line shows the neutral theory prediction, Eq.~\ref{solcontneut}, and the dotted line gives a power-law of exponent $-1$ for comparison. Parameters for (A) are chosen to have large demographic noise: $\nu=0.05$ {\rm day}$^{-1}$, $\mu_0=0.051$ {\rm day}$^{-1}$, $\theta=30$ {\rm day}$^{-1}$ and $p = 0.05$. Parameters for (B) are $\nu=0.05$ {\rm day}$^{-1}$, $\mu_0=0.06$ {\rm day}$^{-1}$, $\theta=40$ {\rm day}$^{-1}$ and $p = 0.07$.}
\label{fig3}
\end{center}
\end{figure}

The stochastic evolution of clone sizes is described by the following Master equation for the mean number $N(C,t)$ of clones of size $C$ at time $t$:
\beq
\begin{split}
\partial_t N(C,t) &=  \nu[(C-1) N(C-1,t) -C N(C,t)] \\
&+ \mu [(C+1) N(C+1,t)-CN(C,t)] + \theta P_0(C).
\end{split}
\eeq
The steady-state distribution of clone sizes can be calculated by solving for $\partial t N(C,t)=0$, yielding for $C>\max\{C: P_0(C)>0\}$,
\beq
N(C) \propto \frac{1}{C}(\nu/\mu)^{C}
\label{neutsol}
\eeq
(see {\em e.g.} Ref.~\cite{desponds2016}, Supplementary Information for a derivation). The distribution falls exponentially fast above the source, creating a large-clone cut-off at $ (\log{\mu/\nu})^{-1}$. Examples of clone size distributions created by this process are shown in Fig.~\ref{fig3} (black curves: analytical prediction; blue curves: simulations).
 
The equation above has a continuous equivalent in the linear-noise approximation, corresponding to the following stochastic differential or Langevin equation:
\beq
\partial_t C_i = (\nu-\mu) C_i + \sqrt{(\mu+\nu)C_i} \xi,
\label{contneut}
\eeq
where $\xi$ is a Gaussian white noise with the \^Ito convention. Note that Eq.~\ref{contneut} was used in Ref.~\cite{Lythe2016} as an approximation of competitive dynamics in the naive immune system to compute the mean lifetime of clones, a case we will discuss further in the next section.
Because $C$ is now continuous, the clone size distribution is described by a probability density $\rho(C,t)$, governed by a Fokker-Planck equation with the steady-state solution \cite{desponds2016}:
\beq
\rho(C) \propto \frac{1}{C}\exp\left( -2\frac{\mu-\nu}{\mu+\nu}C\right).
\label{solcontneut}
\eeq
Although the decay exponents of Eq.~\ref{solcontneut} and Eq.~\ref{neutsol} are different, they agree in the limit where division and death are well balanced, $\mu-\nu \ll \mu+\nu$. This limit is relevant for adaptive immune repertoires (and many neutral systems): it corresponds to division and death rates being tuned to similar values by homeostasis, allowing for large population sizes with minimal thymic output.

The exponential decay of the clone size distribution predicted by the neutral model is inconsistent with the power-law behaviour observed in unsorted T-cell data (see Fig.~\ref{fig1}). However, neutral theory may be consistent with the observed distribution of naive clone cells, although even in this case additional
 mechanisms might be necessary to reproduce data with reasonable parameters \cite{deBoerChain}.

Additionally to neutral dynamics, global competition can also be added to the model to describe the carrying capacity of the population, or homeostasis. Adding this effect does not change the general behaviour of the clone sizes, but simply scales them to ensure a constant population size \cite{desponds2016}. Dividing the population into different groups with different dynamics and differentiation can lead to more complex distributions (see Ref.~\cite{Goyal2015} in the context of hematopoietic stem cell maturation). 
 
 \section{Competition for resources in constant environments}
 \label{cons}
 
The neutral hypothesis has limitations. As already discussed, the clone size distributions it predicts are not consistent with available data. In addition, it ignores competition for survival and division cues in the memory and effector subsets, as well as in the naive subset as shown by experiments in lymphopenic mice \cite{goldrath1999}. 
 
Before we can account for competition, we need to specify the statistics and dynamics of the antigenic landscape, $\{a_j(t)\}$. The temporal dynamics of antigen fluctuations and evolution are generally not well understood. In the simplest assumption, the antigenic landscape is just constant in time, which is plausible for self-antigens, as fluctuations in their concentrations are expected to be spatial rather than temporal and can be averaged over the body. Such a constant landscape of self-antigens seems relevant for naive cells.
On the other hand, the concentrations of the antigens that drive the dynamics of effector and memory cells vary over timescales of days during infections \cite{Fonville2014,Luksza2014,Nourmohammad2015}. 

Lythe et al. \cite{Lythe2016} studied a model of competition of naive T-cells in a constant antigenic landscape. The competitive dynamics are governed by Eq.~\ref{disc}, \ref{stim}, and \ref{ava} with $\epsilon \to \infty$, $\nu(S_i)=\nu_1 S_i$, constant death rate $\mu$, and constant antigenic peptide abundances $a_j(t)=1$. Each element $K_{ij}$ of the interaction matrix is drawn from the same Bernoulli distribution.

Within this model, the mean steady-state repertoire size $T=\<\sum_{i=1}^N C_i\>$ can then be expressed exactly as a function of the biological parameters:
\beq
T=\frac{\nu_1 M+\theta C_0}{\mu},
\label{grant}
\eeq
where we recall that $M$ is the number of antigens, and $\theta C_0$ the thymic output in cells, with $C_0=\<C\>_{P_0}$ the average clone size at introduction. In Ref.~\cite{Lythe2016} the average lifetime of a clone was also {calculated using the neutral approximation of section \ref{neutsec}, i.e. in the limit of negligible competition}. It is concluded from this calculation that the average size of a clone is $10$ in humans, while it is close to $1$ in mice. 

In the general case, there exists no closed equation for $T$. However, progress can be made when  the antigenic peptides abundances $a_j(t)=a$ are uniform and constant, the entries of the interaction matrix $K$ are random and independent, and in the limit where the number of interacting partners of each antigen and of each clone is large. In this limit the availability becomes $F_j\approx F(T)= (1+\epsilon)/(1+\epsilon \<K\>T)$ for all $j$, and the stimulus is $S_i\approx S(T)=Ma \<K\> F(T)$ for all $i$, where $\<K\>$ is the average value of the entries $K_{ij}$. In this case $T$ is given by the implicit equation:
\beq
0=\partial_t T = [ \nu(S(T)) - \mu(S(T)) ] T + \theta C_0.
\eeq

Taking the parameters of the model from de Boer et al. \cite{DeBoer2001} (competition strength $\epsilon=1$, birth rate $\nu=0$, death rate $\mu(S_i)=q/S_i$ {with a constant $q$}, and antigenic peptide abundance $a_j(t)=1/2$) yields:
 \beq
 T \approx \frac{2\<K\>M\theta C_0}{q}\frac{1}{1+\sqrt{1+4\<K\>^2 M\theta C_0/q}}.
 \eeq
Compared to Eq.~\ref{grant}, the population size is given by a geometric rather than arithmetic average between thymic output $\theta C_0$ and antigen pool size $M$ when their product is large, $T\propto \sqrt{M\theta C_0}$.
For comparison, the neutral model predicts a simple linear dependence with thymic output, $T=\theta C_0/(\mu-\nu)$. {With new experiments, these scaling relations could be used to experimentally establish the relevant model classes for particular repertoire subsets.}
 
Going beyond repertoire sizes, one can explore the effect of competition on clone size distributions. Under the assumptions of a constant antigenic environment, there are three sources of fluctuations for clone sizes. The first one is demographic noise, and was already discussed in section \ref{neutsec}. The second one is the fluctuations in the overall ability of clones to bind antigens, measured by $\sum_j K_{ij}$, making some clones intrinsically fitter than others, leading to larger clone sizes. The third source of stochasticity stems from the constant turn-over of clones, which affect the availabilities $F_j$ of antigens, and thus the stimulus $S_i$ that each clone receive. The two last sources of fluctuations could create wider distributions than would be expected from neutral models in certain regimes of parameter space. On the other hand, competition for resources tends to limit the size of the largest clones, which cannot grow beyond the carrying capacity set by antigen availability.

Simulations of Eq,~\ref{gendyn}-\ref{ava}, with added demographic noise within the linear-noise approximation, show that this antigen-availability limit dominates the behaviour of the clone-size distribution for large clones (Fig.~\ref{fig3}). Turning on competition (modeled by the parameter $u>0$ that describes how much the antigen environment affects clone dynamics -- see caption of Fig.~\ref{fig3}) lowers
the large-clone cut-off compared to the neutral theory. The cut-off imposed by resource availability is most visible when demographic noise is important ($\mu - \nu \ll \mu+\nu$, Fig.~\ref{fig3}A).
Like the neutral prediction, these distributions are inconsistent with clone-size distributions observed in effector or memory populations.

 \section{Fluctuating antigenic environments}
 \label{flusec}
 
The dynamics of memory and effector cells are driven by
new pathogens that regularly invade hosts, sometimes triggering full fledged immune responses. Pathogens are then cleared or at least reduced in concentration over rather short time scales. This fast turnover creates a constantly changing antigenic landscape.

Fluctuations in the antigen concentrations keep the system out of equilibrium, with the population of lymphocytes tracking the antigenic landscape with a delay. This situation was studied in Ref.~\cite{desponds2016}, where most clones introduced from thymic or bone marrow output decay exponentially, while a few expand due to strong antigenic stimulus. The clearing of the pathogenic threat depletes the antigenic pool and the expanding clones  shrink back to typical sizes and eventually go extinct. In principle the clearing rate of antigens could vary according to the nature of pathogen or as a function of the efficiency of the immune response. However for simplicity the antigen decay rate $\lambda$ is assumed to be constant, and in this section the effect of competition is ignored, $\epsilon=0$ (we shall relax this assumption in the next section). The division and death rates are set to $\nu(S_i)=\nu_1 S_i$ and constant $\mu$ in Eq.~\ref{gendyn}.

New antigens are introduced with rate $\theta_a$. When a new antigen arrives, some clones experience a transient increase in their effective growth rate $\nu-\mu$, or ``fitness,'' which lasts for a characteristic time $\lambda^{-1}$. Eq.~\ref{gendyn} can be rewritten by separating the constant and fluctuating parts of the fitness as:
\beq
\partial_t C_i(t) = [f_0 +f_i(t)] C_i(t),
\label{dynind}
\eeq
where $f_0=\nu_1\<S_i\>-\mu \leq 0$ gathers the average constant division and death factors and $f_i(t)=\nu_1(S_i-\<S_i\>)$ is the fluctuating part. Although these fluctuations may have complex temporal structure and be correlated between antigens, one can show numerically that they can be well represented by independent Ornstein-Uhlenbeck processes
\beq
\partial_t f_i(t) = -\lambda f_i(t) + \sqrt{2} \gamma \xi_i,
\eeq
where $\xi_i$ is a Gaussian white noise, and
$\gamma$ sets the amplitude of fluctuations, $\<f_i(t)^2\>=\gamma^2/\lambda$. Note that demographic noise in neglected in Eq.~\ref{dynind}, as that noise scales as $\sqrt{C_i}$ and is therefore small compared to fitness fluctuations $\propto C_i$.

In the limit of short $\lambda$, the model is analytically tractable and yields power laws in the clone size distribution, $\rho(C)\propto C^{-1-\alpha}$, consistent with experimental observations. The power law exponent is given by $\alpha=\lambda^2 |f_0|/\gamma^2$: the tail of the distribution is thinner when the decay $|f_0|$ of clones is fast, and heavier when antigenic fluctuations are important (large $\gamma$) and long-lived (small $\lambda$). The power-law behaviour and its exponent are robust to demographic noise, thymic or bone marrow output and the specifics of the interaction matrix $K$.

Experimentally, the observed exponents are very close to $\alpha\approx 1$, as can be seen from Fig.~\ref{fig1}. Plugging in $\alpha=1$, $\lambda=0.1$ {\rm day}$^{-1}$ and $|f_0|=10^{-3}$ {\rm day}$^{-1}$ to the equation for the exponent $\alpha$, we find the typical fluctuations in fitness, $\sqrt{\<f_i(t)^2\>}=\gamma/\sqrt{\lambda}$, are much larger than the average $|f_0|$.

Although this was not done in Ref.~\cite{desponds2016}, the total number of clones in the organism, sometimes simply called ``diversity'' in immunology, or ``species richness'' in ecology, can be estimated from the power-law exponent as $N = T (\alpha-1)/\alpha$ where $T$ is the total number of cells (between $10^{11}$ and $10^{12}$ in humans for T-cells for instance), provided that $\alpha>1$. The average size of a clone is then
$\langle C \rangle = {\alpha}/{(\alpha-1)}$.
This equation depends very sensitively on the value of the exponent around $\alpha=1$. Unfortunately, experimentally observed exponents are often close to $1$, making it hard to get a good estimate. Nonetheless, with the advent of larger datasets and more accurate measurement of the clone-size distribution,
models of fluctuating environments such as the one presented here could help us make predictions for key statistics of the immune system.
 
 \section{Effect of competition on large clones in fluctuating environments}
 The model presented in the previous section ignores competitive effects, making the different clones effectively independent of each other.
What is the effect of explicitly modeling the competition between cells? 

\begin{figure}
\begin{center}
\includegraphics[width=\linewidth]{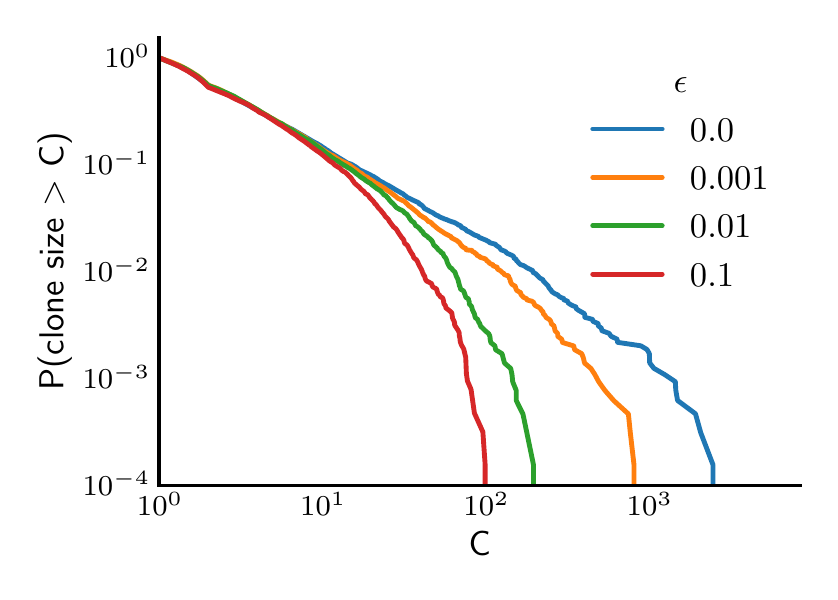}
\caption{Competition affects clone size distribution by limiting the growth of large clones. Cumulative clone size distributions for varying levels of competition after 5000 days. Simulations were performed without demographic noise. A fixed death $\mu=0.003$ day${}^{-1}$, thymic production $\theta = 10$ day${}^{-1}$, and antigen arrival rate $\theta_a = 2$  day${}^{-1}$ were used. Birth rates $\nu(S_i) = \nu_1 S_i$ were rescaled using $\nu_1 = 1.0, 1.08, 1.55, 5.0$ day${}^{-1}$ such that the clone size distributions for different values of $\epsilon$ were similar for small clones. Entries of $K$ were taken to be 1 with probability $p=0.001$ and $0$ otherwise. The size of new clones was set to 2 cells, $P_0(C)=\delta_{C,2}$, the antigen introduction size to $a(t)=1$, and its decay rate to $\lambda = 1$ day${}^{-1}$.}
\label{andreas}
\end{center}
\end{figure}

To study this effect, we simulated the model of Eqs.~\ref{gendyn}-\ref{ava} with $\nu=\nu_1S_i$, constant $\mu$, and a constant source of antigens decaying with rate $\lambda$, as in the previous section, but now with varying levels of competition $\epsilon$.
The resulting steady-state clone-size distributions shown in Fig.~\ref{andreas} show that competition suppresses the power-law behaviour beyond some cut-off in the clone sizes.

Competition can take two forms: interclonal or intraclonal (cells from the same clone competing with each other for antigenic resources). Intuitively, any competition should reduce large clone expansion because of limited resources. 
In particular, intraclonal competition effectively introduces a characteristic clone size, in the form of a carrying capacity, that limits the power-law behaviour. For very large clones $C_i\to\infty$, the growth rate typically scales like the inverse of its size, $\nu\sim C_i^{-1}\nu_1 \sum_{j\in V_i} a_j$, where $V_i$ is the set of antigens that $i$ is specific to,
limiting this size to $C_{i,\rm max}=(\nu_1/\mu)\sum_{j\in V_i} a_j$.
In other words, self-competition prevents clones from growing to populations larger than allowed by antigenic resources.
This result is consistent with the argument of Ref.~\cite{DeBoer2001} that intraclonal competition is essential for naive B-cell homeostasis.

The range of clone size fluctuations can be increased by introducing nonlinearities at various levels: in the dependence of $S_i$ on $a_j$, or in the dependence of the availability function $F_j$ on $C_i$. Such nonlinearities can be justified by the sharp response of T-cells to antigenic affinity and availability \cite{Voisinne2015,Lever2016}. 
 
 \section{Non-specific resources}
 \label{nonspec}
 
So far we have focused on antigenic stimuli as division and survival signals for lymphocytes. However, other growth inducers have been exhibited, particularly cytokines \cite{Schluns2000,Tan2001}. Cytokines do not act through clone-specific receptor binding, so cells from different clones are just as similar or different from each other as cells from the same clone. 
 
If all cells react uniformly to cytokine signaling then the proteins only exercise global homeostasis and the system falls into the class of neutral models, as already discussed in Section~\ref{neutsec}. However, if different cells have access to different levels of cytokine signaling because of variations in the numbers of cytokine receptors, different states of the signaling network, or different locations in the body, more complex dynamics will arise. Mathematically, this heterogeneity can be modeled by adding a cell-specific phenotypic fitness noise reflecting the state and location of the cell.
If this cell state is partially heritable over generations then the cells from the same lineage will have correlated fitness fluctuations. These correlations will decay as the clone grows and information about the ancestral cell state is lost. This class of model gives rise to a wide range of clone size distributions \cite{desponds2016}, some of which are consistent with neutral theory, while others, corresponding to highly heritable cell states, follow an approximate power law over several decades. 

 \section{Aging of immune systems}
 \label{aging}
In all the models above, the system is assumed to reach a steady state over a rather short timescale. These models do not address the question of the long-term effects of competition for resources and homeostasis in immune repertoires. These long-term effects are irrelevant if the antigenic environment fluctuates rapidly, but may be important in stable antigenic landscapes, such as experienced by naive repertoires. The process of competitive exclusion could contribute to depleting repertoire diversity through the selection of ever fitter clones, and could have strong implications for aging of individuals over many years.
 
The evolution of the parameters of lymphocyte homeostasis (e.g. the birth and death rates) has been studied in T-cells in detail \cite{Bains2009,Bains2013a,Hapuarachchi2013}. However the question how whole lymphocyte repertoires change with age remains open. Aging of immune systems entails a loss of diversity at a different pace for different groups of lymphocytes \cite{Goronzy2007} and a general increase in morbidity. Understanding the mechanisms underlying the gradual deterioration of the functioning of the immune system with age, known as immunosenescence, is essential to expanding the efficiency span of vaccines.

The loss of diversity in T-cell repertoires with age could come from different sources: shrinking of the thymus and its output, chronic inflammatory response, overall inefficiency of cell machinery, beneficial somatic mutations driving a few clones to high abundances and depleting others \cite{antia-2012}, or the Hayflick limit \cite{Ndifon2016}. Another possible source is competition, which increases the average affinity of receptors to self-antigens over time, leading to fewer, better-binding receptors \cite{Lythe2016}.
In this section we explore this scenario using our competition model, and study a limit where the increase in binding affinity can be calculated analytically. The model is similar to a classic evolution model of successive selective sweeps \cite{Smith1974}, with the difference that new clones do not originate from existing lineages through mutations, but are produced de novo. The question of how evolution will proceed if driven by the introduction of new clonotypes is mathematically equivalent to evolution driven by mutation in the limit of an infinitely rugged fitness landscape with infinitely many sites, where each mutation leads to a completely random fitness independent of the ancestor \cite{Park2008,Neher2013a}.

\begin{figure} 
\begin{center}
\includegraphics[width=\linewidth]{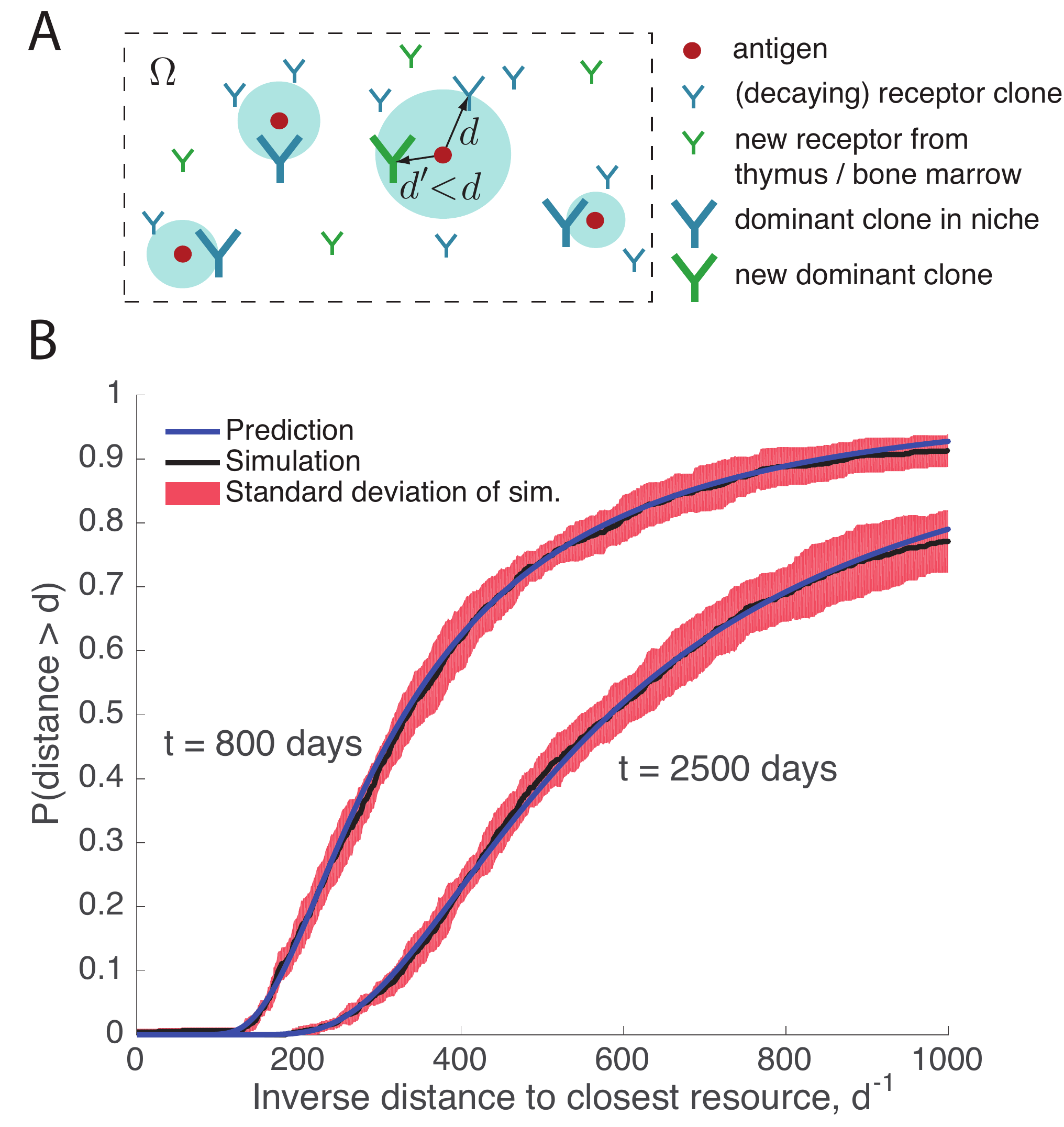}
\caption{Selection of ever more specific clones in a stable antigenic landscape. 
A. Cartoon of the shape space and niche assumption. The fitness of receptor clones in determined by its distance to antigenic resources in the effective shape space $\Omega=(0,1)^d$. Each antigen (red circles) defines an effective niche, in which one clone dominates (large blue receptors), while all others are outcompeted are decay exponentially (small blue receptors). New receptor clones are being introduced from the thymus or bone marrow (small green receptors), and usually stay small or decay because less fit than dominant clones. When a new clones falls inside the blue area, it starts outcompeting the existing dominant clone and displaces it (large green receptor).
B. The dynamics of Eqs.~\ref{gendyn}-\ref{ava} was simulated with $\epsilon=1$, $\nu=1$, and $\mu(S_i)=q/S_i$ with $q=0.01$, and compared to the analytical prediction of Eq.~\ref{record}.
Position of clones and antigens are drawn uniformly at random in an abstract recognition shape space $\Omega=(0,1)^2$, and interact with each other as a function of their distance: $K_{ij}=e^{-d_{ij}/\ell^2}$ with $\ell=0.001$. The number of antigens is $M=1000$. Blue lines are the predictions of Eq.~\ref{record} and black lines are the results of simulations after respectively $800$ and $2500$ days. Parameters are set to $\theta=15$ day${}^{-1}$ and $P_{0}(C)=\delta_{C,2}$.}
\label{selection}
\end{center}
\end{figure}

We model the interaction between clone receptors and antigens as happening in an effective receptor shape space of low dimension. For concreteness we pick this shape space to be a $d$-dimensional hypercube, $\Omega=(0,1)^d$. Both antigens and clones are drawn at random positions uniformly in $\Omega$, as illustrated in Fig.~\ref{selection}A. The interaction strength $K_{ij}$ is determined by the Euclidian distance $d_{ij}$ between the positions of antigen $j$ and clone $i$ in $\Omega$,  $K_{ij}=e^{-d_{ij}^2/\ell^2}$, where $\ell$ is the cross-reactivity range. The antigen pool of size $M$ is drawn at the beginning of the process and kept constant to represent the stability of self-antigens. Clones compete for resources according to Eq.~\ref{ava} with $\epsilon>0$ and their dynamics follow Eq.~\ref{gendyn} with a constant division rate $\nu$ and death rate $\mu(S_i)=q/S_i$. 
 
Each clone has a fitness $g_i=\nu(S_i)-\mu(S_i)$ that depends on its distance to the different resources. It depends strongly on the closest resources because of the fast decay of the binding probability with distance in $\Omega$. Each antigen defines an ecological ``niche'' corresponding to the area of shape space that directly surrounds it: the antigen is the main contributor to the fitness of the clones that fall in that small area. If the spacing of antigens is large compared to the typical interaction range $\ell$, then the different niches become independent: the survival of a given clone depends only on its distance to the closest antigen, as well as on its ability to outcompete other clones depending on the same antigenic resource. 

Under the assumption of competitive exclusion, or strong selection, each niche features one dominant clone that is the closest to the antigen, while all other clones in the niche are outcompeted and decay exponentially fast. As new clones are constantly introduced from thymic and bone marrow output, occasionally a new clone will outcompete the existing dominant clone in a niche and replace it, as schematised in Fig.~\ref{selection}A. The distribution of fitness of dominant clones is simply determined by the fittest clones introduced since the beginning of the process and can be computed analytically.  We call $\Gamma(g,t)$ the probability that the dominant clone in a given niche has fitness smaller than $g$ at time $t$, and $\Gamma_0(g)$ the probability that a newly introduced clone has fitness smaller than $g$, assuming that fitness is dominated by a single antigen (we have dropped the niche index $j$ for notational convenience). Assuming that selective sweeps are fast, the dynamics of the fitness of the dominant clone follow:
\beq
\partial_t \Gamma(g,t)=-\theta[1-\Gamma_0(g)] \Gamma(g,t),
\label{recordeq}
\eeq
where $\theta$ is the rate of introduction of new clones. This equation is solved by:
\beq
\Gamma(g,t) = \Gamma(g,0) e^{-[1-\Gamma_0(g)] \theta t}.\label{record}
\eeq
Eq.~\ref{recordeq} simply states that the clone of largest fitness $g$ is outcompeted at a rate equal to the introduction rate of new clones, $\theta$, multiplied by the probability that a new clone has fitness larger than $g$, $1-\Gamma_0(g)$.
The validity of the dominant-clone approximation can be checked numerically, as shown in Fig.~\ref{selection}B (where the inverse distance $d$ to the antigen is used as a proxy for fitness).

At long times, the process will fall into one of the universality classes of extreme value statistics. More realistic descriptions accounting for demographic noise, or genetic drift, are not discussed here. Despite its simplicity, this basic model shows how the speed of evolution of binding to self-antigens depends on the dimension of the effective shape space, the density of self-antigens relative to the cross-reactivity range $\ell$, and the rate of introduction of new clones.
This approach provides a starting point for analysing the effects of population dynamics on aging immune repertoires.
 
 \section{Conclusion}
 
While a variety of models of evolution of adaptive immune repertoires have been developed \cite{DeBoer1994,DeBoer1995,DeBoer1997,DeBoer2001, Perelson2002, Lythe2016,desponds2016, Johnson2012,Mayer2015, Bains2009,Bains2013a,Hapuarachchi2013}, many of them can be described as different variants of a common set of simple equations (Eqs.~\ref{disc}-\ref{ava}), which are relevant for a variety of repertoire subsets. Specific subclasses of these models (presented in Sections~\ref{flusec} and \ref{nonspec}) predict the observed long-tailed clone size distributions of the combined naive and memory repertoires \cite{Weinstein2009,Mora2010,Zarnitsyna2013,Bolkhovskaya2014a,Menzel2014,Muraro2014,Pogorelyy2016,Mora2016e,deBoerChain} and explain them as a result of a strongly fluctuating environment \cite{desponds2016}. However, while the global features  of the clone size distributions are reproduced, the detailed structure of the competition between cells, which defines how cells experience the fluctuations of the environment, is still not clear. Specifically, the observed distributions are consistent both with clone and cell specific fitness fluctuations. Interestingly, the clone size distributions of naive subsets do not have the long tails of unsorted repertoires, and are in principle consistent with predictions of neutral models presented in Section~\ref{neutsec}  \cite{deBoerChain}. However, more detailed studies are needed to assess the role of competition for self-antigens also in these subsets. Lastly, the same theoretical considerations can be extended to study the open question of the aging of immune repertoires. In Section~\ref{aging} we proposed a model of aging based on selective sweeps of dominant clonoytpes in independent niches. 
With the advent of repertoire data from organisms of different ages \cite{Britanova2014,Britanova2016}, it will be interesting to verify these models and see how the collective population dynamics of lymphocytes changes with age.   
 
We have shown here that using stochastic population level models can offer insights into lymphocyte dynamics from static measurements of the clone size distributions. These models have great interpretive and predictive power, however whole repertoire data is still relatively rare and clone size distributions only provide us with one way of probing the dynamics. The idea of using a snapshot of the immune repertoire to learn about its history can be extended beyond distributions of clone sizes by using the other features extracted from receptor sequence distributions that can provide clearer insight into the underlying dynamics. For example, how similar receptor sequences are in phenotypic space is a potential marker of both competitive exclusion and co-variation of co-stimulated clones. This task requires solving the very hard question of linking distances in phenotypic space with sequences, or simultaneously probing the phenotypic space of many receptors in high throughput experiments. Experiments going in these directions are currently being developed \cite{Adams2016, Boyer2016a}. Alternatively, one could look for appropriate metrics directly in sequence space that inform us about competitive exclusion and the distribution of clonotypes \cite{chain-2014, Epstein2014}. In both cases,  performing such anaylyses requires overcoming both the data sampling problem and building the bioinformatics tools to recover these distributions from data.

The adaptive immune system is of course vastly more complex than suggested by the models presented in this paper. In particular, the relative importance of naive and memory repertoires with age, thymic output, division and death rates of cells and the overall decay with age of different functions of the body could have unexpected effects on repertoire statistics and should be further investigated. The adaptability of immune systems implies that they depend strongly on their history and we can expect that certain properties of the repertoire will show history dependence. For example, experiments suggest the existence of long-lived clones that benefited from less competitive homeostatis conditions in utero to expand to large sizes \cite{Pogorelyy2016}. 

The mathematical models discussed here pay little attention to the initial conditions of the population dynamics, as specified by early repertoire maturation and  development. This simplification is motivated by the quick establishment of a steady state, or at least of an adiabatically changing steady state. However, some characteristics of repertoire development in utero can survive for decades \cite{Pogorelyy2016}. An interesting direction would be include the effect of these early maturation events in population dynamics models. Such extensions could help to explain the inequality of susceptibility to infection and morbidity across individuals in a given species.


\begin{thebibliography}{66}
\expandafter\ifx\csname natexlab\endcsname\relax\def\natexlab#1{#1}\fi
\expandafter\ifx\csname bibnamefont\endcsname\relax
  \def\bibnamefont#1{#1}\fi
\expandafter\ifx\csname bibfnamefont\endcsname\relax
  \def\bibfnamefont#1{#1}\fi
\expandafter\ifx\csname citenamefont\endcsname\relax
  \def\citenamefont#1{#1}\fi
\expandafter\ifx\csname url\endcsname\relax
  \def\url#1{\texttt{#1}}\fi
\expandafter\ifx\csname urlprefix\endcsname\relax\def\urlprefix{URL }\fi
\providecommand{\bibinfo}[2]{#2}
\providecommand{\eprint}[2][]{\url{#2}}

\bibitem[{\citenamefont{Janeway}(2005)}]{Janeway}
\bibinfo{author}{\bibfnamefont{C.}~\bibnamefont{Janeway}},
  \emph{\bibinfo{title}{Immunobiology}} (\bibinfo{publisher}{Garland Science},
  \bibinfo{year}{2005}).

\bibitem[{\citenamefont{Troy and Shen}(2003)}]{Troy2003}
\bibinfo{author}{\bibfnamefont{A.~E.} \bibnamefont{Troy}} \bibnamefont{and}
  \bibinfo{author}{\bibfnamefont{H.}~\bibnamefont{Shen}}, \bibinfo{journal}{J.
  Immunol.} \textbf{\bibinfo{volume}{170}} (\bibinfo{year}{2003}).

\bibitem[{\citenamefont{Mak and Saunders}(2006)}]{mak2006immune}
\bibinfo{author}{\bibfnamefont{T.}~\bibnamefont{Mak}} \bibnamefont{and}
  \bibinfo{author}{\bibfnamefont{M.}~\bibnamefont{Saunders}},
  \emph{\bibinfo{title}{The Immune Response: Basic and Clinical Principles}},
  \bibinfo{number}{vol.~1} (\bibinfo{publisher}{Elsevier/Academic},
  \bibinfo{year}{2006}).

\bibitem[{\citenamefont{Weinstein et~al.}(2009)\citenamefont{Weinstein, Jiang,
  White, Fisher, and Quake}}]{Weinstein2009}
\bibinfo{author}{\bibfnamefont{J.~A.} \bibnamefont{Weinstein}},
  \bibinfo{author}{\bibfnamefont{N.}~\bibnamefont{Jiang}},
  \bibinfo{author}{\bibfnamefont{R.~A.} \bibnamefont{White}},
  \bibinfo{author}{\bibfnamefont{D.~S.} \bibnamefont{Fisher}},
  \bibnamefont{and} \bibinfo{author}{\bibfnamefont{S.~R.} \bibnamefont{Quake}},
  \bibinfo{journal}{Science (80-. ).} \textbf{\bibinfo{volume}{324}},
  \bibinfo{pages}{807} (\bibinfo{year}{2009}).

\bibitem[{\citenamefont{Boyd et~al.}(2009)}]{Boyd2009a}
\bibinfo{author}{\bibfnamefont{S.~D.} \bibnamefont{Boyd}} \bibnamefont{et~al.},
  \bibinfo{journal}{Sci Transl Med} \textbf{\bibinfo{volume}{1}}
  (\bibinfo{year}{2009}).

\bibitem[{\citenamefont{Robins et~al.}(2009)}]{Robins2009}
\bibinfo{author}{\bibfnamefont{H.~S.} \bibnamefont{Robins}}
  \bibnamefont{et~al.}, \bibinfo{journal}{Blood}
  \textbf{\bibinfo{volume}{114}}, \bibinfo{pages}{4099} (\bibinfo{year}{2009}).

\bibitem[{\citenamefont{Ndifon et~al.}(2012)}]{friedman-2012}
\bibinfo{author}{\bibfnamefont{W.}~\bibnamefont{Ndifon}} \bibnamefont{et~al.},
  \bibinfo{journal}{Proceedings of the National Academy of Sciences}
  \textbf{\bibinfo{volume}{109}}, \bibinfo{pages}{15865}
  (\bibinfo{year}{2012}).

\bibitem[{\citenamefont{Thomas et~al.}(2014)}]{chain-2014}
\bibinfo{author}{\bibfnamefont{N.}~\bibnamefont{Thomas}} \bibnamefont{et~al.},
  \bibinfo{journal}{Bioinformatics} \textbf{\bibinfo{volume}{30}},
  \bibinfo{pages}{3181} (\bibinfo{year}{2014}).

\bibitem[{\citenamefont{Larimore et~al.}(2012)\citenamefont{Larimore,
  McCormick, Robins, and Greenberg}}]{greenberg-2012}
\bibinfo{author}{\bibfnamefont{K.}~\bibnamefont{Larimore}},
  \bibinfo{author}{\bibfnamefont{M.~W.} \bibnamefont{McCormick}},
  \bibinfo{author}{\bibfnamefont{H.~S.} \bibnamefont{Robins}},
  \bibnamefont{and} \bibinfo{author}{\bibfnamefont{P.~D.}
  \bibnamefont{Greenberg}}, \bibinfo{journal}{J Immunol}
  (\bibinfo{year}{2012}).

\bibitem[{\citenamefont{Sherwood et~al.}(2011)}]{robins-2011}
\bibinfo{author}{\bibfnamefont{A.~M.} \bibnamefont{Sherwood}}
  \bibnamefont{et~al.}, \bibinfo{journal}{Sci Transl Med}
  \textbf{\bibinfo{volume}{3}} (\bibinfo{year}{2011}).

\bibitem[{\citenamefont{Zvyagin et~al.}(2014)}]{mamedov-2014}
\bibinfo{author}{\bibfnamefont{I.~V.} \bibnamefont{Zvyagin}}
  \bibnamefont{et~al.}, \bibinfo{journal}{Proceedings of the National Academy
  of Sciences}  (\bibinfo{year}{2014}).

\bibitem[{\citenamefont{Warren et~al.}(2011)}]{Warren2011}
\bibinfo{author}{\bibfnamefont{R.~L.} \bibnamefont{Warren}}
  \bibnamefont{et~al.}, \bibinfo{journal}{Genome Research}
  \textbf{\bibinfo{volume}{21}}, \bibinfo{pages}{790} (\bibinfo{year}{2011}).

\bibitem[{\citenamefont{Murugan et~al.}(2012)\citenamefont{Murugan, Mora,
  Walczak, and Callan}}]{murugan-2012}
\bibinfo{author}{\bibfnamefont{A.}~\bibnamefont{Murugan}},
  \bibinfo{author}{\bibfnamefont{T.}~\bibnamefont{Mora}},
  \bibinfo{author}{\bibfnamefont{A.~M.} \bibnamefont{Walczak}},
  \bibnamefont{and} \bibinfo{author}{\bibfnamefont{C.~G.}
  \bibnamefont{Callan}}, \bibinfo{journal}{Proceedings of the National Academy
  of Sciences} \textbf{\bibinfo{volume}{109}}, \bibinfo{pages}{16161}
  (\bibinfo{year}{2012}).

\bibitem[{\citenamefont{Elhanati et~al.}(2014)\citenamefont{Elhanati, Murugan,
  Callan~Jr, Mora, and Walczak}}]{walczak-2014}
\bibinfo{author}{\bibfnamefont{Y.}~\bibnamefont{Elhanati}},
  \bibinfo{author}{\bibfnamefont{A.}~\bibnamefont{Murugan}},
  \bibinfo{author}{\bibfnamefont{C.~G.} \bibnamefont{Callan~Jr}},
  \bibinfo{author}{\bibfnamefont{T.}~\bibnamefont{Mora}}, \bibnamefont{and}
  \bibinfo{author}{\bibfnamefont{A.~M.} \bibnamefont{Walczak}},
  \bibinfo{journal}{Proceedings of the National Academy of Sciences}
  \textbf{\bibinfo{volume}{111}}, \bibinfo{pages}{9875} (\bibinfo{year}{2014}).

\bibitem[{\citenamefont{Yaari et~al.}(2013)}]{Yaari2013a}
\bibinfo{author}{\bibfnamefont{G.}~\bibnamefont{Yaari}} \bibnamefont{et~al.},
  \bibinfo{journal}{Front. Immunol.} \textbf{\bibinfo{volume}{4}},
  \bibinfo{pages}{358} (\bibinfo{year}{2013}).

\bibitem[{\citenamefont{Elhanati et~al.}(2015)}]{Elhanati2015}
\bibinfo{author}{\bibfnamefont{Y.}~\bibnamefont{Elhanati}}
  \bibnamefont{et~al.}, \bibinfo{journal}{Philos Trans R Soc Lond, B, Biol Sci}
  \textbf{\bibinfo{volume}{370}}, \bibinfo{pages}{20140243}
  (\bibinfo{year}{2015}).

\bibitem[{\citenamefont{Mccoy et~al.}(2015)}]{Mccoy2015}
\bibinfo{author}{\bibfnamefont{C.~O.} \bibnamefont{Mccoy}}
  \bibnamefont{et~al.}, \bibinfo{journal}{Philos Trans R Soc Lond, B, Biol Sci}
  \textbf{\bibinfo{volume}{370}}, \bibinfo{pages}{20140244}
  (\bibinfo{year}{2015}).

\bibitem[{\citenamefont{Vollmers et~al.}(2013)\citenamefont{Vollmers, Sit,
  Weinstein, Dekker, and Quake}}]{Vollmers2013}
\bibinfo{author}{\bibfnamefont{C.}~\bibnamefont{Vollmers}},
  \bibinfo{author}{\bibfnamefont{R.~V.} \bibnamefont{Sit}},
  \bibinfo{author}{\bibfnamefont{J.~A.} \bibnamefont{Weinstein}},
  \bibinfo{author}{\bibfnamefont{C.~L.} \bibnamefont{Dekker}},
  \bibnamefont{and} \bibinfo{author}{\bibfnamefont{S.~R.} \bibnamefont{Quake}},
  \bibinfo{journal}{Proceedings of the National Academy of Sciences}
  \textbf{\bibinfo{volume}{110}}, \bibinfo{pages}{13463}
  (\bibinfo{year}{2013}).

\bibitem[{\citenamefont{Best et~al.}(2015)\citenamefont{Best, Oakes, Heather,
  Shawe-Taylor, and Chain}}]{Best2015b}
\bibinfo{author}{\bibfnamefont{K.}~\bibnamefont{Best}},
  \bibinfo{author}{\bibfnamefont{T.}~\bibnamefont{Oakes}},
  \bibinfo{author}{\bibfnamefont{J.~M.} \bibnamefont{Heather}},
  \bibinfo{author}{\bibfnamefont{J.}~\bibnamefont{Shawe-Taylor}},
  \bibnamefont{and} \bibinfo{author}{\bibfnamefont{B.}~\bibnamefont{Chain}},
  \bibinfo{journal}{Sci. Rep.} \textbf{\bibinfo{volume}{5}},
  \bibinfo{pages}{14629} (\bibinfo{year}{2015}).

\bibitem[{\citenamefont{Shugay et~al.}(2014)}]{Shugay2014a}
\bibinfo{author}{\bibfnamefont{M.}~\bibnamefont{Shugay}} \bibnamefont{et~al.},
  \bibinfo{journal}{Nat. Methods} \textbf{\bibinfo{volume}{11}},
  \bibinfo{pages}{653} (\bibinfo{year}{2014}).

\bibitem[{\citenamefont{Laserson et~al.}(2014)}]{Laserson2014}
\bibinfo{author}{\bibfnamefont{U.}~\bibnamefont{Laserson}}
  \bibnamefont{et~al.}, \bibinfo{journal}{Proc. Natl. Acad. Sci.}
  \textbf{\bibinfo{volume}{111}}, \bibinfo{pages}{4928} (\bibinfo{year}{2014}).

\bibitem[{\citenamefont{Galson et~al.}(2014)\citenamefont{Galson, Pollard,
  Tr{\"{u}}ck, and Kelly}}]{Galson2014}
\bibinfo{author}{\bibfnamefont{J.~D.} \bibnamefont{Galson}},
  \bibinfo{author}{\bibfnamefont{A.~J.} \bibnamefont{Pollard}},
  \bibinfo{author}{\bibfnamefont{J.}~\bibnamefont{Tr{\"{u}}ck}},
  \bibnamefont{and} \bibinfo{author}{\bibfnamefont{D.~F.} \bibnamefont{Kelly}},
  \bibinfo{journal}{Trends Immunol.} \textbf{\bibinfo{volume}{35}},
  \bibinfo{pages}{319} (\bibinfo{year}{2014}).

\bibitem[{\citenamefont{Wu et~al.}(2012)}]{Wu2012}
\bibinfo{author}{\bibfnamefont{D.}~\bibnamefont{Wu}} \bibnamefont{et~al.},
  \bibinfo{journal}{Sci. Transl. Med.} \textbf{\bibinfo{volume}{4}}
  (\bibinfo{year}{2012}).

\bibitem[{\citenamefont{Salson et~al.}(2016)}]{Salson2016}
\bibinfo{author}{\bibfnamefont{M.}~\bibnamefont{Salson}} \bibnamefont{et~al.},
  \bibinfo{journal}{Leuk. Res.} \textbf{\bibinfo{volume}{53}},
  \bibinfo{pages}{1} (\bibinfo{year}{2016}).

\bibitem[{\citenamefont{Mora et~al.}(2010)\citenamefont{Mora, Walczak, Bialek,
  and Callan}}]{Mora2010}
\bibinfo{author}{\bibfnamefont{T.}~\bibnamefont{Mora}},
  \bibinfo{author}{\bibfnamefont{A.~M.} \bibnamefont{Walczak}},
  \bibinfo{author}{\bibfnamefont{W.}~\bibnamefont{Bialek}}, \bibnamefont{and}
  \bibinfo{author}{\bibfnamefont{C.~G.} \bibnamefont{Callan}},
  \bibinfo{journal}{Proceedings of the National Academy of Sciences}
  \textbf{\bibinfo{volume}{107}}, \bibinfo{pages}{5405} (\bibinfo{year}{2010}).

\bibitem[{\citenamefont{Zarnitsyna et~al.}(2013)\citenamefont{Zarnitsyna,
  Evavold, Schoettle, Blattman, and Antia}}]{Zarnitsyna2013}
\bibinfo{author}{\bibfnamefont{V.~I.} \bibnamefont{Zarnitsyna}},
  \bibinfo{author}{\bibfnamefont{B.~D.} \bibnamefont{Evavold}},
  \bibinfo{author}{\bibfnamefont{L.~N.} \bibnamefont{Schoettle}},
  \bibinfo{author}{\bibfnamefont{J.~N.} \bibnamefont{Blattman}},
  \bibnamefont{and} \bibinfo{author}{\bibfnamefont{R.}~\bibnamefont{Antia}},
  \bibinfo{journal}{Front. Immunol.} \textbf{\bibinfo{volume}{4}},
  \bibinfo{pages}{485} (\bibinfo{year}{2013}).

\bibitem[{\citenamefont{Bolkhovskaya et~al.}(2014)\citenamefont{Bolkhovskaya,
  Zorin, and Ivanchenko}}]{Bolkhovskaya2014a}
\bibinfo{author}{\bibfnamefont{O.~V.} \bibnamefont{Bolkhovskaya}},
  \bibinfo{author}{\bibfnamefont{D.~Y.} \bibnamefont{Zorin}}, \bibnamefont{and}
  \bibinfo{author}{\bibfnamefont{M.~V.} \bibnamefont{Ivanchenko}},
  \bibinfo{journal}{PLoS One} \textbf{\bibinfo{volume}{9}},
  \bibinfo{pages}{e108658} (\bibinfo{year}{2014}).

\bibitem[{\citenamefont{Menzel et~al.}(2014)\citenamefont{Menzel, Greiff, Khan,
  Haessler, Hellmann, Friedensohn, Cook, Pogson, and Reddy}}]{Menzel2014}
\bibinfo{author}{\bibfnamefont{U.}~\bibnamefont{Menzel}},
  \bibinfo{author}{\bibfnamefont{V.}~\bibnamefont{Greiff}},
  \bibinfo{author}{\bibfnamefont{T.~A.} \bibnamefont{Khan}},
  \bibinfo{author}{\bibfnamefont{U.}~\bibnamefont{Haessler}},
  \bibinfo{author}{\bibfnamefont{I.}~\bibnamefont{Hellmann}},
  \bibinfo{author}{\bibfnamefont{S.}~\bibnamefont{Friedensohn}},
  \bibinfo{author}{\bibfnamefont{S.~C.} \bibnamefont{Cook}},
  \bibinfo{author}{\bibfnamefont{M.}~\bibnamefont{Pogson}}, \bibnamefont{and}
  \bibinfo{author}{\bibfnamefont{S.~T.} \bibnamefont{Reddy}},
  \bibinfo{journal}{PLoS One} \textbf{\bibinfo{volume}{9}}, \bibinfo{pages}{1}
  (\bibinfo{year}{2014}).

\bibitem[{\citenamefont{Muraro and Robins}(2014)}]{Muraro2014}
\bibinfo{author}{\bibfnamefont{P.}~\bibnamefont{Muraro}} \bibnamefont{and}
  \bibinfo{author}{\bibfnamefont{H.}~\bibnamefont{Robins}},
  \bibinfo{journal}{J. {\ldots}} \textbf{\bibinfo{volume}{124}},
  \bibinfo{pages}{1168} (\bibinfo{year}{2014}).

\bibitem[{\citenamefont{Pogorelyy et~al.}(2016)\citenamefont{Pogorelyy,
  Elhanati, Marcou, Sycheva, Komech, Nazarov, Britanova, Chudakov, Mamedov,
  Lebedev et~al.}}]{Pogorelyy2016}
\bibinfo{author}{\bibfnamefont{M.~V.} \bibnamefont{Pogorelyy}},
  \bibinfo{author}{\bibfnamefont{Y.}~\bibnamefont{Elhanati}},
  \bibinfo{author}{\bibfnamefont{Q.}~\bibnamefont{Marcou}},
  \bibinfo{author}{\bibfnamefont{A.~L.} \bibnamefont{Sycheva}},
  \bibinfo{author}{\bibfnamefont{E.~A.} \bibnamefont{Komech}},
  \bibinfo{author}{\bibfnamefont{V.~I.} \bibnamefont{Nazarov}},
  \bibinfo{author}{\bibfnamefont{O.~V.} \bibnamefont{Britanova}},
  \bibinfo{author}{\bibfnamefont{D.~M.} \bibnamefont{Chudakov}},
  \bibinfo{author}{\bibfnamefont{I.~Z.} \bibnamefont{Mamedov}},
  \bibinfo{author}{\bibfnamefont{Y.~B.} \bibnamefont{Lebedev}},
  \bibnamefont{et~al.}, \bibinfo{journal}{arXiv} p. \bibinfo{pages}{1602.03063}
  (\bibinfo{year}{2016}).

\bibitem[{\citenamefont{Mora and Walczak}(2016)}]{Mora2016e}
\bibinfo{author}{\bibfnamefont{T.}~\bibnamefont{Mora}} \bibnamefont{and}
  \bibinfo{author}{\bibfnamefont{A.}~\bibnamefont{Walczak}},
  \bibinfo{journal}{ArXiv} p. \bibinfo{pages}{1604.00487}
  (\bibinfo{year}{2016}).

\bibitem[{\citenamefont{de~Greef et~al.}(2017)\citenamefont{de~Greef, Oakes,
  Gerritsen, Heather, Hermsen, and de~Boer}}]{deBoerChain}
\bibinfo{author}{\bibfnamefont{P.~C.} \bibnamefont{de~Greef}},
  \bibinfo{author}{\bibfnamefont{T.}~\bibnamefont{Oakes}},
  \bibinfo{author}{\bibfnamefont{B.}~\bibnamefont{Gerritsen}},
  \bibinfo{author}{\bibfnamefont{J.~M.} \bibnamefont{Heather}},
  \bibinfo{author}{\bibfnamefont{B.}~\bibnamefont{Hermsen},
  \bibfnamefont{Rutger~Chain}}, \bibnamefont{and}
  \bibinfo{author}{\bibfnamefont{R.~J.} \bibnamefont{de~Boer}},
  \bibinfo{journal}{In preparation}  (\bibinfo{year}{2017}).

\bibitem[{\citenamefont{Perelson}(2002)}]{Perelson2002}
\bibinfo{author}{\bibfnamefont{A.~S.} \bibnamefont{Perelson}},
  \bibinfo{journal}{Nat. Rev. Immunol.} \textbf{\bibinfo{volume}{2}},
  \bibinfo{pages}{28} (\bibinfo{year}{2002}).

\bibitem[{\citenamefont{Johnson
  et~al.}(2012{\natexlab{a}})\citenamefont{Johnson, Yates, Goronzy, and
  Antia}}]{Johnson2012}
\bibinfo{author}{\bibfnamefont{P.~L.~F.} \bibnamefont{Johnson}},
  \bibinfo{author}{\bibfnamefont{A.~J.} \bibnamefont{Yates}},
  \bibinfo{author}{\bibfnamefont{J.~J.} \bibnamefont{Goronzy}},
  \bibnamefont{and} \bibinfo{author}{\bibfnamefont{R.}~\bibnamefont{Antia}},
  \bibinfo{journal}{Proceedings of the National Academy of Sciences}
  \textbf{\bibinfo{volume}{109}}, \bibinfo{pages}{21432}
  (\bibinfo{year}{2012}{\natexlab{a}}).

\bibitem[{\citenamefont{Mayer et~al.}(2015)\citenamefont{Mayer,
  Balasubramanian, Mora, and Walczak}}]{Mayer2015}
\bibinfo{author}{\bibfnamefont{A.}~\bibnamefont{Mayer}},
  \bibinfo{author}{\bibfnamefont{V.}~\bibnamefont{Balasubramanian}},
  \bibinfo{author}{\bibfnamefont{T.}~\bibnamefont{Mora}}, \bibnamefont{and}
  \bibinfo{author}{\bibfnamefont{A.~M.} \bibnamefont{Walczak}},
  \bibinfo{journal}{Proc. Natl. Acad. Sci.} \textbf{\bibinfo{volume}{112}},
  \bibinfo{pages}{5950} (\bibinfo{year}{2015}).

\bibitem[{\citenamefont{Lythe et~al.}(2016)\citenamefont{Lythe, Callard, Hoare,
  and Molina-par{\'{\i}}s}}]{Lythe2016}
\bibinfo{author}{\bibfnamefont{G.}~\bibnamefont{Lythe}},
  \bibinfo{author}{\bibfnamefont{R.~E.} \bibnamefont{Callard}},
  \bibinfo{author}{\bibfnamefont{R.~L.} \bibnamefont{Hoare}}, \bibnamefont{and}
  \bibinfo{author}{\bibfnamefont{C.}~\bibnamefont{Molina-par{\'{\i}}s}},
  \bibinfo{journal}{Journal of Theoretical Biology}
  \textbf{\bibinfo{volume}{389}}, \bibinfo{pages}{214} (\bibinfo{year}{2016}).

\bibitem[{\citenamefont{Kimura}(1983)}]{kimurabook}
\bibinfo{author}{\bibfnamefont{M.}~\bibnamefont{Kimura}},
  \emph{\bibinfo{title}{The Neutral Theory of Molecular Evolution}}
  (\bibinfo{publisher}{Cambridge University Press}, \bibinfo{year}{1983}).

\bibitem[{\citenamefont{Nielsen}(2005)}]{Nielsen2005}
\bibinfo{author}{\bibfnamefont{R.}~\bibnamefont{Nielsen}},
  \bibinfo{journal}{Annu. Rev. Genet} \textbf{\bibinfo{volume}{39}},
  \bibinfo{pages}{197} (\bibinfo{year}{2005}).

\bibitem[{\citenamefont{{De Boer} and Perelson}(1994)}]{DeBoer1994}
\bibinfo{author}{\bibfnamefont{R.~J.} \bibnamefont{{De Boer}}}
  \bibnamefont{and} \bibinfo{author}{\bibfnamefont{A.~S.}
  \bibnamefont{Perelson}}, \bibinfo{journal}{Journal of theoretical biology}
  \textbf{\bibinfo{volume}{169}}, \bibinfo{pages}{375} (\bibinfo{year}{1994}).

\bibitem[{\citenamefont{{De Boer} and Perelson}(1995)}]{DeBoer1995}
\bibinfo{author}{\bibfnamefont{R.~J.} \bibnamefont{{De Boer}}}
  \bibnamefont{and} \bibinfo{author}{\bibfnamefont{a.~S.}
  \bibnamefont{Perelson}}, \bibinfo{journal}{J. Theor. Biol.}
  \textbf{\bibinfo{volume}{175}}, \bibinfo{pages}{567} (\bibinfo{year}{1995}).

\bibitem[{\citenamefont{{De Boer} and Perelson}(1997)}]{DeBoer1997}
\bibinfo{author}{\bibfnamefont{R.~J.} \bibnamefont{{De Boer}}}
  \bibnamefont{and} \bibinfo{author}{\bibfnamefont{A.~S.}
  \bibnamefont{Perelson}}, \bibinfo{journal}{Int. Immunol.}
  \textbf{\bibinfo{volume}{9}}, \bibinfo{pages}{779} (\bibinfo{year}{1997}).

\bibitem[{\citenamefont{{De Boer} et~al.}(2001)\citenamefont{{De Boer},
  Freitas, and Perelson}}]{DeBoer2001}
\bibinfo{author}{\bibfnamefont{R.~J.} \bibnamefont{{De Boer}}},
  \bibinfo{author}{\bibfnamefont{A.~A.} \bibnamefont{Freitas}},
  \bibnamefont{and} \bibinfo{author}{\bibfnamefont{A.~S.}
  \bibnamefont{Perelson}}, \bibinfo{journal}{Journal of theoretical biology}
  \textbf{\bibinfo{volume}{212}}, \bibinfo{pages}{333} (\bibinfo{year}{2001}).

\bibitem[{\citenamefont{Desponds et~al.}(2016)\citenamefont{Desponds, Mora, and
  Walczak}}]{desponds2016}
\bibinfo{author}{\bibfnamefont{J.}~\bibnamefont{Desponds}},
  \bibinfo{author}{\bibfnamefont{T.}~\bibnamefont{Mora}}, \bibnamefont{and}
  \bibinfo{author}{\bibfnamefont{A.~M.} \bibnamefont{Walczak}},
  \bibinfo{journal}{Proceedings of the National Academy of Sciences}
  \textbf{\bibinfo{volume}{113}}, \bibinfo{pages}{274} (\bibinfo{year}{2016}).

\bibitem[{\citenamefont{Goyal et~al.}(2015)\citenamefont{Goyal, Kim, Chen, and
  Chou}}]{Goyal2015}
\bibinfo{author}{\bibfnamefont{S.}~\bibnamefont{Goyal}},
  \bibinfo{author}{\bibfnamefont{S.}~\bibnamefont{Kim}},
  \bibinfo{author}{\bibfnamefont{I.~S.} \bibnamefont{Chen}}, \bibnamefont{and}
  \bibinfo{author}{\bibfnamefont{T.}~\bibnamefont{Chou}}, \bibinfo{journal}{BMC
  Biology} \textbf{\bibinfo{volume}{13}}, \bibinfo{pages}{85}
  (\bibinfo{year}{2015}).

\bibitem[{\citenamefont{Goldrath and Bevan}(1999)}]{goldrath1999}
\bibinfo{author}{\bibfnamefont{A.~W.} \bibnamefont{Goldrath}} \bibnamefont{and}
  \bibinfo{author}{\bibfnamefont{M.~J.} \bibnamefont{Bevan}},
  \bibinfo{journal}{Immunity} \textbf{\bibinfo{volume}{11}},
  \bibinfo{pages}{183} (\bibinfo{year}{1999}).

\bibitem[{\citenamefont{Fonville et~al.}(2014)}]{Fonville2014}
\bibinfo{author}{\bibfnamefont{J.~M.} \bibnamefont{Fonville}}
  \bibnamefont{et~al.}, \bibinfo{journal}{Science}
  \textbf{\bibinfo{volume}{346}}, \bibinfo{pages}{7} (\bibinfo{year}{2014}).

\bibitem[{\citenamefont{Luksza and L{\"{a}}ssig}(2014)}]{Luksza2014}
\bibinfo{author}{\bibfnamefont{M.}~\bibnamefont{Luksza}} \bibnamefont{and}
  \bibinfo{author}{\bibfnamefont{M.}~\bibnamefont{L{\"{a}}ssig}},
  \bibinfo{journal}{Nature} \textbf{\bibinfo{volume}{507}}, \bibinfo{pages}{57}
  (\bibinfo{year}{2014}).

\bibitem[{\citenamefont{Nourmohammad et~al.}(2016)\citenamefont{Nourmohammad,
  Otwinowski, and Plotkin}}]{Nourmohammad2015}
\bibinfo{author}{\bibfnamefont{A.}~\bibnamefont{Nourmohammad}},
  \bibinfo{author}{\bibfnamefont{J.}~\bibnamefont{Otwinowski}},
  \bibnamefont{and} \bibinfo{author}{\bibfnamefont{J.~B.}
  \bibnamefont{Plotkin}}, \bibinfo{journal}{PloS Genet}
  \textbf{\bibinfo{volume}{12}} (\bibinfo{year}{2016}).

\bibitem[{\citenamefont{Voisinne et~al.}(2015)\citenamefont{Voisinne, Nixon,
  and Vergassola}}]{Voisinne2015}
\bibinfo{author}{\bibfnamefont{G.}~\bibnamefont{Voisinne}},
  \bibinfo{author}{\bibfnamefont{B.~G.} \bibnamefont{Nixon}}, \bibnamefont{and}
  \bibinfo{author}{\bibfnamefont{M.}~\bibnamefont{Vergassola}},
  \bibinfo{journal}{Cells Reports} pp. \bibinfo{pages}{1208--1219}
  (\bibinfo{year}{2015}).

\bibitem[{\citenamefont{Lever et~al.}(2016)}]{Lever2016}
\bibinfo{author}{\bibfnamefont{M.}~\bibnamefont{Lever}} \bibnamefont{et~al.},
  \bibinfo{journal}{Proceedings of the National Academy of Sciences}
  (\bibinfo{year}{2016}).

\bibitem[{\citenamefont{Schluns et~al.}(2000)\citenamefont{Schluns, Kieper,
  Jameson, and Lefran\c{c}ois}}]{Schluns2000}
\bibinfo{author}{\bibfnamefont{K.~S.} \bibnamefont{Schluns}},
  \bibinfo{author}{\bibfnamefont{W.~C.} \bibnamefont{Kieper}},
  \bibinfo{author}{\bibfnamefont{S.~C.} \bibnamefont{Jameson}},
  \bibnamefont{and}
  \bibinfo{author}{\bibfnamefont{L.}~\bibnamefont{Lefran\c{c}ois}},
  \bibinfo{journal}{Nature immunology} \textbf{\bibinfo{volume}{1}},
  \bibinfo{pages}{426} (\bibinfo{year}{2000}).

\bibitem[{\citenamefont{Tan et~al.}(2001)}]{Tan2001}
\bibinfo{author}{\bibfnamefont{J.~T.} \bibnamefont{Tan}} \bibnamefont{et~al.},
  \bibinfo{journal}{Proceedings of the National Academy of Sciences}
  \textbf{\bibinfo{volume}{98}}, \bibinfo{pages}{8732} (\bibinfo{year}{2001}).

\bibitem[{\citenamefont{Bains et~al.}(2009)\citenamefont{Bains, Antia, Callard,
  and Yates}}]{Bains2009}
\bibinfo{author}{\bibfnamefont{I.}~\bibnamefont{Bains}},
  \bibinfo{author}{\bibfnamefont{R.}~\bibnamefont{Antia}},
  \bibinfo{author}{\bibfnamefont{R.}~\bibnamefont{Callard}}, \bibnamefont{and}
  \bibinfo{author}{\bibfnamefont{A.~J.} \bibnamefont{Yates}},
  \bibinfo{journal}{{Blood}} \textbf{\bibinfo{volume}{113}},
  \bibinfo{pages}{5480} (\bibinfo{year}{2009}).

\bibitem[{\citenamefont{Bains et~al.}(2013)\citenamefont{Bains, Yates, and
  Callard}}]{Bains2013a}
\bibinfo{author}{\bibfnamefont{I.}~\bibnamefont{Bains}},
  \bibinfo{author}{\bibfnamefont{A.~J.} \bibnamefont{Yates}}, \bibnamefont{and}
  \bibinfo{author}{\bibfnamefont{R.~E.} \bibnamefont{Callard}},
  \bibinfo{journal}{PLoS One} \textbf{\bibinfo{volume}{8}}
  (\bibinfo{year}{2013}).

\bibitem[{\citenamefont{Hapuarachchi et~al.}(2013)\citenamefont{Hapuarachchi,
  Lewis, and Callard}}]{Hapuarachchi2013}
\bibinfo{author}{\bibfnamefont{T.}~\bibnamefont{Hapuarachchi}},
  \bibinfo{author}{\bibfnamefont{J.}~\bibnamefont{Lewis}}, \bibnamefont{and}
  \bibinfo{author}{\bibfnamefont{R.~E.} \bibnamefont{Callard}},
  \bibinfo{journal}{Frontiers in Immunology} \textbf{\bibinfo{volume}{4}},
  \bibinfo{pages}{2} (\bibinfo{year}{2013}).

\bibitem[{\citenamefont{Goronzy et~al.}(2007)\citenamefont{Goronzy, Lee, and
  Weyand}}]{Goronzy2007}
\bibinfo{author}{\bibfnamefont{J.~J.} \bibnamefont{Goronzy}},
  \bibinfo{author}{\bibfnamefont{W.~W.} \bibnamefont{Lee}}, \bibnamefont{and}
  \bibinfo{author}{\bibfnamefont{C.~M.} \bibnamefont{Weyand}},
  \bibinfo{journal}{Exp. Gerontol.} \textbf{\bibinfo{volume}{42}},
  \bibinfo{pages}{400} (\bibinfo{year}{2007}).

\bibitem[{\citenamefont{Johnson
  et~al.}(2012{\natexlab{b}})\citenamefont{Johnson, Yates, Goronzy, and
  Antia}}]{antia-2012}
\bibinfo{author}{\bibfnamefont{P.~L.~F.} \bibnamefont{Johnson}},
  \bibinfo{author}{\bibfnamefont{A.~J.} \bibnamefont{Yates}},
  \bibinfo{author}{\bibfnamefont{J.~J.} \bibnamefont{Goronzy}},
  \bibnamefont{and} \bibinfo{author}{\bibfnamefont{R.}~\bibnamefont{Antia}},
  \bibinfo{journal}{Proc. Natl. Acad. Sci.} \textbf{\bibinfo{volume}{109}},
  \bibinfo{pages}{21432} (\bibinfo{year}{2012}{\natexlab{b}}).

\bibitem[{\citenamefont{Ndifon and Dushoff}(2016)}]{Ndifon2016}
\bibinfo{author}{\bibfnamefont{W.}~\bibnamefont{Ndifon}} \bibnamefont{and}
  \bibinfo{author}{\bibfnamefont{J.}~\bibnamefont{Dushoff}},
  \bibinfo{journal}{J. Immunol.}  (\bibinfo{year}{2016}).

\bibitem[{\citenamefont{Smith and Haigh}(1974)}]{Smith1974}
\bibinfo{author}{\bibfnamefont{J.~M.} \bibnamefont{Smith}} \bibnamefont{and}
  \bibinfo{author}{\bibfnamefont{J.}~\bibnamefont{Haigh}},
  \bibinfo{journal}{Genetical research} \textbf{\bibinfo{volume}{23}}
  (\bibinfo{year}{1974}).

\bibitem[{\citenamefont{Park and Krug}(2008)}]{Park2008}
\bibinfo{author}{\bibfnamefont{S.-c.} \bibnamefont{Park}} \bibnamefont{and}
  \bibinfo{author}{\bibfnamefont{J.}~\bibnamefont{Krug}},
  \bibinfo{journal}{Journal of Statistical mechanics}  (\bibinfo{year}{2008}).

\bibitem[{\citenamefont{Neher et~al.}(2013)\citenamefont{Neher, Vucelja,
  Mezard, and Shraiman}}]{Neher2013a}
\bibinfo{author}{\bibfnamefont{R.~A.} \bibnamefont{Neher}},
  \bibinfo{author}{\bibfnamefont{M.}~\bibnamefont{Vucelja}},
  \bibinfo{author}{\bibfnamefont{M.}~\bibnamefont{Mezard}}, \bibnamefont{and}
  \bibinfo{author}{\bibfnamefont{B.~I.} \bibnamefont{Shraiman}},
  \bibinfo{journal}{J. Stat. Mech. Theory Exp.}
  \textbf{\bibinfo{volume}{2013}}, \bibinfo{pages}{P01008}
  (\bibinfo{year}{2013}).

\bibitem[{\citenamefont{Britanova et~al.}(2014)\citenamefont{Britanova,
  Putintseva, Shugay, Merzlyak, Turchaninova, Staroverov, Bolotin, Lukyanov,
  Bogdanova, Mamedov et~al.}}]{Britanova2014}
\bibinfo{author}{\bibfnamefont{O.~V.} \bibnamefont{Britanova}},
  \bibinfo{author}{\bibfnamefont{E.~V.} \bibnamefont{Putintseva}},
  \bibinfo{author}{\bibfnamefont{M.}~\bibnamefont{Shugay}},
  \bibinfo{author}{\bibfnamefont{E.~M.} \bibnamefont{Merzlyak}},
  \bibinfo{author}{\bibfnamefont{M.~A.} \bibnamefont{Turchaninova}},
  \bibinfo{author}{\bibfnamefont{D.~B.} \bibnamefont{Staroverov}},
  \bibinfo{author}{\bibfnamefont{D.~A.} \bibnamefont{Bolotin}},
  \bibinfo{author}{\bibfnamefont{S.}~\bibnamefont{Lukyanov}},
  \bibinfo{author}{\bibfnamefont{E.~A.} \bibnamefont{Bogdanova}},
  \bibinfo{author}{\bibfnamefont{I.~Z.} \bibnamefont{Mamedov}},
  \bibnamefont{et~al.}, \bibinfo{journal}{J. Immunol.}
  \textbf{\bibinfo{volume}{192}}, \bibinfo{pages}{2689} (\bibinfo{year}{2014}).

\bibitem[{\citenamefont{Britanova et~al.}(2016)\citenamefont{Britanova, Shugay,
  Merzlyak, Staroverov, Putintseva, Turchaninova, Mamedov, Pogorelyy, Bolotin,
  Izraelson et~al.}}]{Britanova2016}
\bibinfo{author}{\bibfnamefont{O.~V.} \bibnamefont{Britanova}},
  \bibinfo{author}{\bibfnamefont{M.}~\bibnamefont{Shugay}},
  \bibinfo{author}{\bibfnamefont{E.~M.} \bibnamefont{Merzlyak}},
  \bibinfo{author}{\bibfnamefont{D.~B.} \bibnamefont{Staroverov}},
  \bibinfo{author}{\bibfnamefont{E.~V.} \bibnamefont{Putintseva}},
  \bibinfo{author}{\bibfnamefont{M.~A.} \bibnamefont{Turchaninova}},
  \bibinfo{author}{\bibfnamefont{I.~Z.} \bibnamefont{Mamedov}},
  \bibinfo{author}{\bibfnamefont{M.~V.} \bibnamefont{Pogorelyy}},
  \bibinfo{author}{\bibfnamefont{D.~A.} \bibnamefont{Bolotin}},
  \bibinfo{author}{\bibfnamefont{M.}~\bibnamefont{Izraelson}},
  \bibnamefont{et~al.}, \bibinfo{journal}{J. Immunol.}  (\bibinfo{year}{2016}).

\bibitem[{\citenamefont{Adams et~al.}(2016)\citenamefont{Adams, Kinney, Mora,
  and Walczak}}]{Adams2016}
\bibinfo{author}{\bibfnamefont{R.~M.} \bibnamefont{Adams}},
  \bibinfo{author}{\bibfnamefont{J.~B.} \bibnamefont{Kinney}},
  \bibinfo{author}{\bibfnamefont{T.}~\bibnamefont{Mora}}, \bibnamefont{and}
  \bibinfo{author}{\bibfnamefont{A.~M.} \bibnamefont{Walczak}},
  \bibinfo{journal}{Elife} p. \bibinfo{pages}{1601.02160}
  (\bibinfo{year}{2016}).

\bibitem[{\citenamefont{Boyer et~al.}(2016)\citenamefont{Boyer, Biswas, {Kumar
  Soshee}, Scaramozzino, Nizak, and Rivoire}}]{Boyer2016a}
\bibinfo{author}{\bibfnamefont{S.}~\bibnamefont{Boyer}},
  \bibinfo{author}{\bibfnamefont{D.}~\bibnamefont{Biswas}},
  \bibinfo{author}{\bibfnamefont{A.}~\bibnamefont{{Kumar Soshee}}},
  \bibinfo{author}{\bibfnamefont{N.}~\bibnamefont{Scaramozzino}},
  \bibinfo{author}{\bibfnamefont{C.}~\bibnamefont{Nizak}}, \bibnamefont{and}
  \bibinfo{author}{\bibfnamefont{O.}~\bibnamefont{Rivoire}},
  \bibinfo{journal}{Proc. Natl. Acad. Sci.} \textbf{\bibinfo{volume}{113}},
  \bibinfo{pages}{3482} (\bibinfo{year}{2016}).

\bibitem[{\citenamefont{Epstein et~al.}(2014)\citenamefont{Epstein, Barenco,
  Klein, Hubank, and Callard}}]{Epstein2014}
\bibinfo{author}{\bibfnamefont{M.}~\bibnamefont{Epstein}},
  \bibinfo{author}{\bibfnamefont{M.}~\bibnamefont{Barenco}},
  \bibinfo{author}{\bibfnamefont{N.}~\bibnamefont{Klein}},
  \bibinfo{author}{\bibfnamefont{M.}~\bibnamefont{Hubank}}, \bibnamefont{and}
  \bibinfo{author}{\bibfnamefont{R.~E.} \bibnamefont{Callard}},
  \bibinfo{journal}{PLoS One} \textbf{\bibinfo{volume}{9}}
  (\bibinfo{year}{2014}).

\end{thebibliography}
\end{document}